\documentclass[useAMS,usenatbib]{mn2e}
\usepackage{graphicx}
\usepackage{multirow}
\usepackage{amssymb}

\title[Complex asteroseismology of $\nu$ Eridani]
      {Complex asteroseismology of the $\beta$ Cep/SPB pulsator \\ $\nu$ Eridani: constraints on opacities}

   \author[J. Daszy\'nska-Daszkiewicz and P. Walczak]
{J. Daszy\'nska-Daszkiewicz\thanks{E-mail:daszynska@astro.uni.wroc.pl(JDD); walczak@astro.uni.wroc.pl(PW)}
and P. Walczak\footnotemark[1]\\
   Instytut Astronomiczny, Uniwersytet Wroc{\l}awski,
   ul. Kopernika 11, 51-622 Wroc{\l}aw, Poland\\
    }
    \begin{document}

   \date{Received ...; accepted ...:in original form ...}

   \pagerange{\pageref{firstpage}--\pageref{lastpage}} \pubyear{2009}

   \maketitle

\label{firstpage}

\begin{abstract}
We undertake another attempt towards seismic modelling of the most intensive
studied main sequence pulsators of the early B spectral type, $\nu$ Eridani.
Our analysis is extended by a requirement of fitting not only pulsational frequencies
but also the complex amplitude of the bolometric flux variation, $f$, related
to each mode frequency. This approach, called {\it complex asteroseismology},
provides a unique test of stellar parameters, atmospheres and opacities.
In particular, the concordance of the empirical and theoretical values of $f$
we obtained for the high-order g mode opens a new gate in seismic studies
of the main-sequence hybrid pulsators.
The most intriguing and challenging result is that whereas an agreement
of the theoretical and empirical values of $f$ for the radial mode can be achieved
only with the OPAL data, a preference for the OP tables is derived from
the analysis of the high-order gravity mode.
\end{abstract}


\section{Introduction}
Oscillation frequencies are the primary data used in seismic modelling.
If oscillation spectra are rich and some regularities can be identified,
then a substantial amount of information can be extracted about stellar interior and physics.
This is the case for the Sun, solar-like stars and some white dwarfs,
for which equidistances in frequencies or periods are observed.

In recent years, we have entered the age of studying pulsating variables
from space, by missions such as MOST, CoRoT, Kelper, which are delivering
great and interesting results. But despite the tremendous number
of detected pulsational frequencies, there is no main sequence pulsator
showing regular structures in its oscillation spectrum.
The best example is a $\delta$ Scuti star, HD50844, for which
at least 1000 frequencies were detected from the CoRoT photometry
and no clear regularity in frequency spacing is seen (Poretti et al. 2009).
Moreover, all satellite missions observe in one wide photometric passband,
therefore nothing can be said about mode identification and
a support from ground-based observations is unavoidable.
A hope for asterseismology of main-sequence stars
can be put in the forthcoming BRITE mission which will make
a two-passband photometry supported by ground-based spectroscopy.

Therefore, for the time being, an improvement in seismic modelling
of main-sequence pulsators can be rather achieved by incorporating
other seismic tools related to each mode frequency.
One of such tools is the relative amplitude of radiative flux perturbation
at the level of the photosphere, which is called the $f$-parameter.
Because of the non-adiabaticity of pulsation this quantity is complex.
The $f$-parameter is determined in subphotospheric layers,
mainly in the pulsation driving region, whereas oscillation frequencies
are defined mainly by the stellar interior.
Therefore these two seismic probes are complementary to each other
and fitting them simultaneously, which we called {\it complex asteroseismology}
(Daszy\'nska-Daszkiewicz \& Walczak 2009, hereafter DDW09),
can bring much stronger constraints on parameters of the model and theory.
Theoretical values of $f$ are obtained in the framework of linear nonadiabatic
computations of stellar pulsation and their empirical counterparts
are derived from multicolour time series photometry.

Asteroseismic potential of the $f$-parameter has been already proven for some
$\delta$ Scuti stars (Daszy\'nska-Daszkiewicz, Dziembowski, Pamyatnykh, 2003,
Daszy\'nska-Daszkiewicz et al. 2005a) and $\beta$ Cephei stars
(Daszy\'nska-Daszkiewicz, Dziembowski, Pamyatnykh 2005, hereafter, DD05).
In the case of $\delta$ Scuti stars, the $f$-parameter appeared to be strongly dependent
on the treatment of subphotospheric convection. For the B-type pulsators,
a strong sensitivity to the metal content and the opacities has been found.
Thus, a comparison of empirical and theoretical values of $f$ can yield information
on the convective transport in the cooler pulsators and on opacities in the hotter ones.

In the present paper, we make one more attempt towards seismic modelling of $\nu$ Eridani,
which is a B-type hybrid pulsator, i.e., one showing both $\beta$ Cephei type pulsations
and frequencies typical for Slowly Pulsating B-type (SPB) stars.
In our study we aim at parallel fitting of pulsation frequencies and the
corresponding values of the complex, nonadiabatic parameter $f$, taking into account
instability conditions. Such studies have been done recently by us for the $\beta$ Cep star
$\theta$ Ophiuchi (DDW09). From fitting two centroid frequencies of $\theta$ Oph
and the $f$-parameter of the radial mode, we have found a strong preference
for the OPAL opacity tables.
In the case of $\nu$ Eri, we have more useable frequencies, so that
stronger constraints can be expected. Our main seismic modelling is based
on fitting three p-mode frequencies. Then, we single out models which fit
the forth p-mode frequency and two low frequencies corresponding to high-order g modes.
Moreover, empirical values of $f$ can be determined in a wide range of frequencies,
i.e., from  about 0.6 to 8 c/d.
Another important result of the presented paper is identification of the mode
degree, $\ell$, for all frequencies detected in the recent photometric multi-site
campaign (Jerzykiewicz et al. 2005, hereafter J05) using two methods.

In Section 2, we give a brief review of $\nu$ Eridani.
Section 3 is devoted to identifying of the spherical harmonic degree, $\ell$,
for all fourteen modes using two approaches. A discussion of empirical values
of the intrinsic mode-amplitude is also included.
In Section 4, we present results of our seismic modelling of $\nu$ Eri from fitting
pulsational frequencies and then from reconciling corresponding values of
the complex, nonadiabatic parameter $f$. Conclusions end the paper.

\section{$\nu$ Eri and its pulsational modes}
$\nu$ Eridani (HD 29248, $V=3.92$ mag) is the most extensively studied main-sequence pulsator
of early B spectral type (B2III). The star is a slow rotator, with the velocity of rotation of about 6 km/s
derived from the rotational splitting of pulsational frequencies (Pamyatnykh, Handler \& Dziembowski, 2004).
During the last few years, this variable has attracted much interest thanks to dedicated photometric
(Handler et al. 2004, J05) and spectroscopic (Aerts et al. 2004) multisite campaigns.
These observations allowed to detect new frequencies typical for $\beta$ Cep stars
as well as entirely new two peaks in the SPB frequency domain.
The frequency analysis revealed 14 pulsational frequencies in photometry (Handler et al. 2004, J05), while 9 of them,
including one SPB frequency, were identified also in spectroscopy (Aerts et al. 2004).
As a result, $\nu$  Eridani has became one of the most multiperiodic B-type main-sequence
pulsator and the updated oscillation spectrum (shown in Fig.\,1) required to reclassify it
from the $\beta$ Cep to $\beta$ Cep/SPB type. Recently, another $\beta$ Cep/SPB star
rivals it in the number of pulsation frequencies. From an analysis of MOST photometry
of $\gamma$ Peg, Handler et al. (2009) found 14 independent peaks: 8 of the $\beta$ Cep type
and 6 of the SPB type.

Mode identification of the $\nu$ Eri frequencies from photometric
and spectroscopic observations (Ridder at el. 2004) and from inspection of the oscillation
spectrum (J05) showed the well known radial, fundamental mode and three $\ell=1$ dipole modes,
$\rm{g_1},~\rm{p_1}$ and $\rm{p_2}$, with all components. Identification of the remaining modes,
including two SPB frequencies, was not successful or has not been made.

Subsequently, several papers were devoted to seismic analysis of $\nu$ Eri.
Pamyatnykh, Handler \& Dziembowski (2004) encountered a problem with
excitation of the high-order g mode, $\nu_A=0.4328$ c/d, as well as the high-frequency p modes
with $\nu> 6$ c/d. As a solution, these authors proposed increasing the iron abundance
in the driving region; this excited the high frequency modes but not the g mode.
From an analysis of the rotational splitting of two dipole modes, they deduced that
rotation of $\nu$ Eri is not uniform, with its rate in the $\mu$-gradient zone
about three times faster than in the outer layers. Pamyatnykh, Handler \& Dziembowski (2004)
managed to find models which reproduced the three centroid frequencies corresponding to the
$\ell=0,\rm{p_1}$, $\ell=1,\rm{g_1}$ and $\ell=1,\rm{p_1}$ modes but did not succeed in fitting
the highest centroid frequency, a component of the $\ell=1,\rm{p_2}$ triplet.
This was done by Ausseloos et al. (2004) but for higher values of the overshooting parameter,
$\alpha_{\rm ov}\approx 0.3$, effective temperatures much lower than $3\sigma$ of
the observational value or non-standard chemical composition.
Moreover, the problem with the $\ell=1,\rm{p_2}$ mode excitation
for models with a standard chemical composition remained.
Both papers relied on one source of opacity data, OPAL, and on
the old solar chemical composition (Grevesse \& Noels, 1993).
\begin{figure}
\centering
\includegraphics[width=85mm,clip]{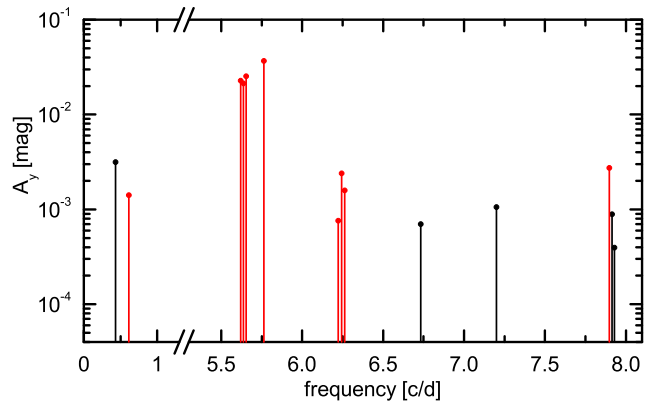}
\caption{ The oscillation spectrum of $\nu$ Eridani. Fourteen
frequencies were detected from photometry and
nine of them were found also in spectroscopic variations,
marked as red peaks. The frequency region $1.2-5.2$ c/d, in which no frequencies were detected,
is not shown.} \label{aaaaaa}
\end{figure}
\begin{figure*}
\centering
\includegraphics[width=17.8cm,clip]{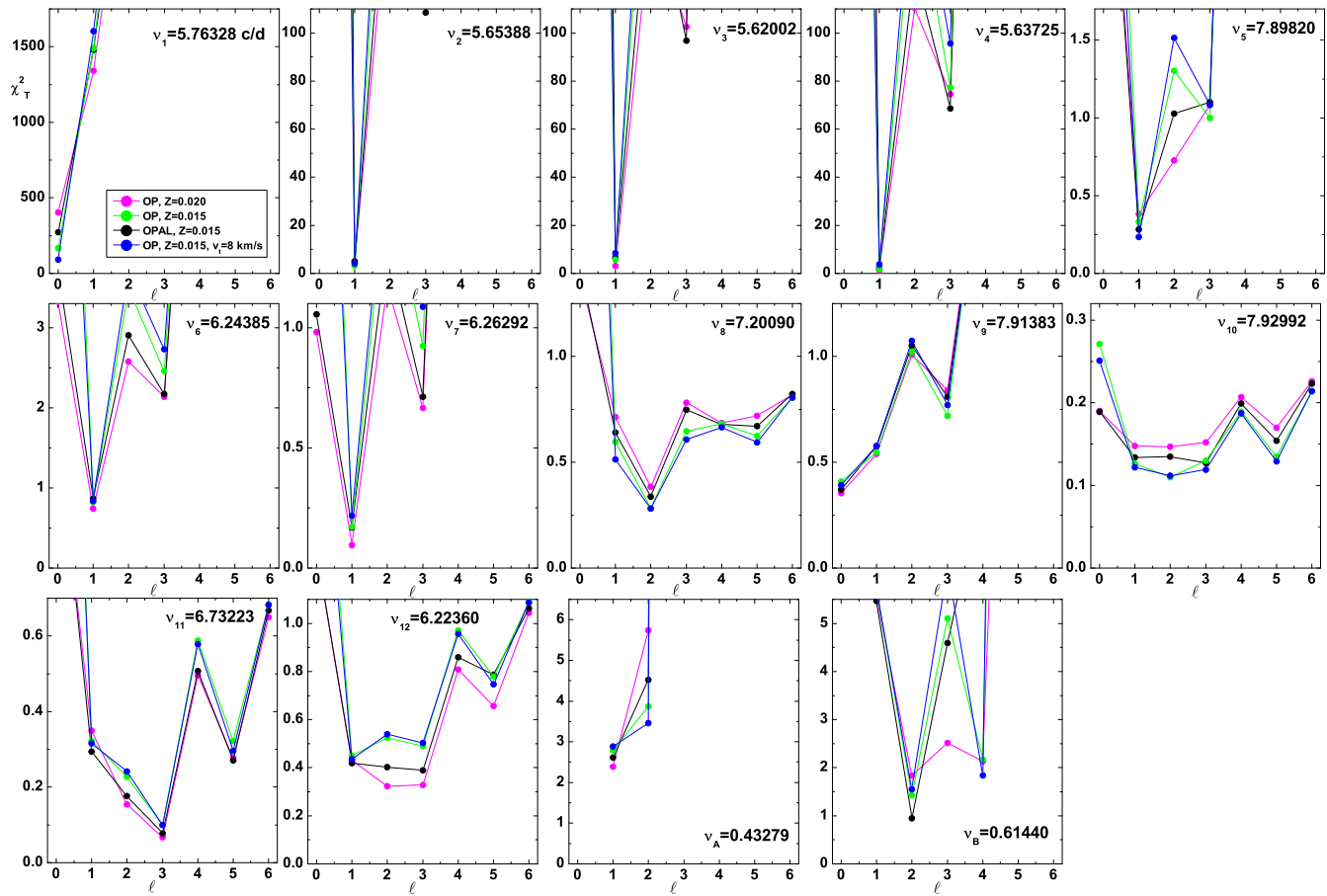}
\caption{ The $\chi^2_T$ discriminant as a function of $\ell$,
obtained from fitting the photometric amplitude ratios and phase differences
for 14 frequencies of $\nu$ Eri detected in the Str\"omgren $uvy$ light-curves.
Results are shown for four stellar models with the mass of 9.5 $M_\odot$,
metallicity, $Z$=0.02 and 0.015, OP and OPAL opacities, and the microturbulent
velocity in the atmosphere, $\xi_t=2$ and 8 km/s.}
\label{aaaaaa}
\end{figure*}

DD05 undertook the first attempt to fit also the $f$-parameter for the radial mode
using both OPAL and OP tables.
They showed that, in the case of B-type pulsators, $f$ strongly depends
on metallicity and the adopted opacity data. A comparison of empirical values of $f$
with theoretical ones corresponding to four seismic models with $\alpha_{\rm ov}=0.0$ and 0.1
showed that no agreement is possible regardless of which opacity tables are used.

More recently, seismic studies were done by Dziembowski \& Pamyatnykh (2008), These authors
used OPAL and OP data and included a new solar composition (Asplund et al. 2004, hereafter A04).
In that paper, Dziembowski \& Pamyatnykh (2008) abandoned the idea of iron accumulation
in the $Z-$bump zone to explain the g mode excitation, because there is no evidence
of an abundance anomaly in the atmosphere of $\nu$ Eri.

Effects of differential rotation in the analysis of three $\ell=1$ triplets
of $\nu$ Eri were included by Suarez et al. (2009).

Detection of high-order g-mode pulsation in massive stars ($M>7M_{\odot}$)
creates new challenges and opportunities for theory of stellar pulsation.
Firstly, a problem of mode excitation in this frequency range
for massive models should be solved. This can be achieved by improving
opacity computations and more precise seismic modelling which
should include, e.g., chemical composition of a given star.
Secondly, multi-colour photometric data on modes allow to determine
empirical values of $f$ in a wide frequency range and to compare them
with theoretical counterparts. This can provide unique constraints
because a dependence of the $f$-parameter on the mode frequency and degree, $\ell$,
is very different for p and g modes.

\section{Identification of the degree $\ell$ of pulsational modes}
Although several years have elapsed since the second photometric campaign of $\nu$ Eridani
(J05), mode identification employing photometric observables for all 14 detected pulsational
peaks has never been undertaken.

The $uvy$ Str\"omgran photometry and radial velocity measurements allowed us to apply
two approaches to identify the degree $\ell$ of the pulsational modes.
In the first case, we compare observational values of the amplitude ratios and phase differences
between available passpands with their theoretical counterparts and rely on theoretical
values of the $f$-parameter which result from linear nonadiabatic computations
of stellar oscillations (Cugier, Dziembowski \& Pamyatnykh, 1994).
In the second method we make use of the amplitudes and phases themselves
and a value of $f$ is determined in the Least Square process together with $\ell$.
For a detailed description of this method see Daszy\'nska-Daszkiewicz, Dziembowski \& Pamyatnykh (2003) and DD05.
In the case of B-type pulsators, the second method demands radial velocity measurements
to get a unique identification of the degree $\ell$.
As for input from stellar atmospheres, we use Kurucz (2004) models
and Claret's limb darkening coefficients (Claret 2000).
Both methods provide a discriminant as a function of $\ell$, which we named $\chi^2_T$
or $\chi^2_E$, according to whether theoretical or empirical values of $f$ were used.
For definitions see DDW09.
\begin{figure*}
\centering
\includegraphics[width=17.8cm,clip]{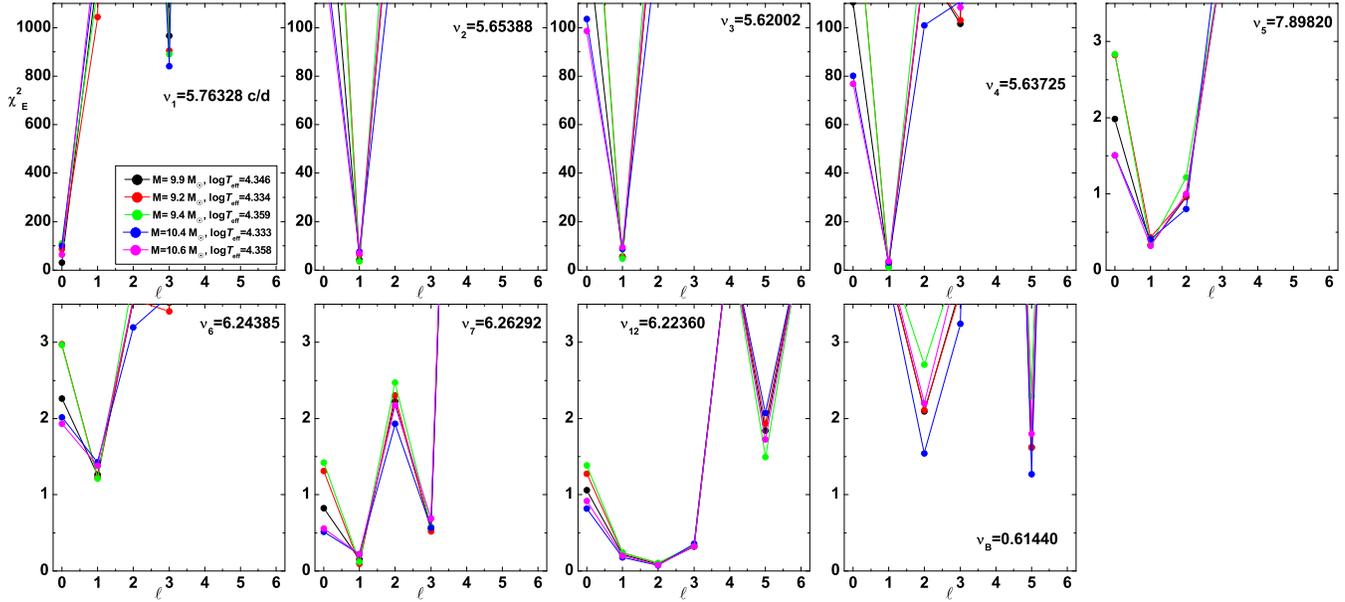}
\caption{Values of $\chi^2_E$ as a function of $\ell$ determined from fitting amplitudes
and phases of the light and radial velocity variations for nine frequencies detected
in both photometric and spectroscopic observations of $\nu$ Eridani.
Results are shown for five models: from the centre and the four edges of the error box (Eq.\,1).}
\label{aaaaaa}
\end{figure*}

In this paper, we assume observational values of the effective temperature and luminosity
derived by DD05, i.e.
$$\log T_{\rm eff}= 4.346\pm 0.014 ;~~  \log L/L_{\odot} =3.94\pm 0.16, \eqno(1)$$
To determine the values of $\chi_T^2$ for the fourteen frequencies detected
in photometric variations of $\nu$ Eri, we used models with various masses,
luminosities, effective temperatures and metallicites within the error box,
assuming the A04 chemical mixture.
As an example, in Fig.\,2 we plot $\chi_T^2$ as a function of the mode degree, $\ell$,
determined for four models with the same mass, $M=9.5 M_\odot$, but with two values of metallicity,
$Z=0.020$ and 0.015, two opacity tables, OPAL and OP, and two values of the microturbulent velocity
in the atmosphere, $\xi_t=2$ and 8 km/s.

It is well established that the dominant frequency, $\nu_1$, is the radial fundamental mode.
The three dipole modes with all components, ($\nu_2$, $\nu_3$, $\nu_4$), ($\nu_6$, $\nu_7$, $\nu_{12}$)
and ($\nu_5$, $\nu_9$, $\nu_{10}$), were identified in earlier papers either from photometric observables
(Ridder et al. 2004) or from the oscillation spectrum itself (J05).
As one can see from Fig.\,2, these identifications are confirmed by the method
of amplitude ratios and phase differences. Identification of the degree $\ell$
for $\nu_8$, $\nu_{11}$ and for two SPB modes, $\nu_A$ and $\nu_B$, is as follows.
For $\nu_8=7.2009$ c/d, $\chi^2_T$ reaches low values at all $\ell$, except $\ell=0$,
but there is some indication that $\ell=2$. A similar result is obtained for $\nu_{11}=6.7322$ c/d,
but here $\chi_T^2$ has a minimum at $\ell=3$.
We have to add in advance that whereas $\nu_8$ matches $\ell=2$ in seismic pulsational models,
the frequency $\nu_{11}$ can be associated only with $\ell=4$. The same identifications for these two
frequencies were proposed by Dziembowski \& Pamyatnykh (2008).
Luckily, an unambiguous identification of $\ell$ was possible for the high-order g modes:
$\nu_A=0.433$ c/d is an $\ell=1$ mode, whereas $\nu_B=0.614$ c/d is an $\ell=2$ mode.
For the latter, there is some probability of $\ell=4$, but it is much smaller
because visibility of this mode is much worse.

Now, let us discuss in more detail the values of $\chi_T^2$ for $\nu_1$, the radial mode.
Firstly, these values are much higher than 1 whichever model is considered. This is because
formal errors of the photometric amplitudes and phases are underestimated (J05);
this problem has been already discussed by DD05.
Another, more relevant feature is that a value of $\chi^2_T(\nu_1)$ decreases about
twofold if the microturbulent velocity of 8 km/s is adopted instead of 2 km/s,
with other parameters fixed. This result agrees with determination of $\xi_t$ in the atmosphere
of $\nu$ Eri by Gies \& Lambert (1992) who obtained $\xi_t=9.4\pm 2.6$ within the LTE approximation.

The second method of mode identification can be applied for nine frequencies,
including one SPB freqyency, which were found in multicolour photometry and in spectroscopy.
From line profile variations we derived the values of radial velocity, defined as the first moment
of the SiIII4553{\AA} spectral line. In Fig.\,3, we show the $\chi^2_E$ dependence
on $\ell$ for five models: for the centre and the four edges of the error box (Eq. 1).
The $\ell$ identifications for first seven frequencies, $\nu_1-\nu_7$, and the twelfth, $\nu_{12}$,
agree with the previous ones. Now, let us deal with the low frequency mode, $\nu_B=0.614$ c/d.
As can be seen from Fig.\,3, the discriminant $\chi^2_E$ reaches a minimum for $\ell=2$  and $\ell=5$.
This approach excludes $\ell=4$, whereas the first method excluded $\ell=5$.
Moreover, these higher $\ell$ modes suffer from strong disc-averaging effects.
Thus, we regard the $\ell=2$ identification as most probable.
A summary of our identifications of $\ell$ by the two methods for
all 14 pulsational frequencies of $\nu$ Eri is given in Tab.\,1.
\begin{table}
\begin{center}
\caption{Most probable identification of the $\ell$ degree
for the pulsational frequencies of $\nu$ Eri from two approaches.}
\begin{tabular}{|l|c|c|}
\hline
\multirow{2}{*}{frequency [c/d]}& photometry       & phot.+$V_{\rm{rad}}$\\
                             & (theoretical $f$'s) & (empirical $f$'s)
\\ \hline
$\nu_{1}$	=	5.7632828	&	$\ell$=0	&	$\ell$=0	\\
$\nu_{2}$	=	5.6538767	&	$\ell$=1	&	$\ell$=1	\\
$\nu_{3}$	=	5.6200186	&	$\ell$=1	&	$\ell$=1	\\
$\nu_{4}$	=	5.6372470	&	$\ell$=1	&	$\ell$=1	\\
$\nu_{5}$	=	7.898200	&	$\ell$=1	&	$\ell$=1	\\
$\nu_{6}$	=	6.243847	&	$\ell$=1	&	$\ell$=1	\\
$\nu_{7}$	=	6.262917	&	$\ell$=1,3  &	$\ell$=1,3	\\
$\nu_{8}$	=	7.20090	    &	$\ell$=?(2)	&	-	\\
$\nu_{9}$	=	7.91383	    &	$\ell\le 3$	&	-	\\
$\nu_{10}$	=	7.92992 	&	$\ell$=?	&	-	\\
$\nu_{11}$	=	6.73223 	&	$\ell$=?(3)	&	-	\\
$\nu_{12}$	=	6.22360	    &  $\ell$=1,2,3 &	$\ell$=1,2,3	\\
$\nu_{A}$	=	0.432786	&	$\ell$=1	&	-	\\
$\nu_{B}$	=	0.61440	    &	$\ell$=2,4	&	$\ell$=2,5	\\
\hline
\end{tabular}
\label{tab_mse}
\end{center}
\end{table}

Other quantities extracted with the second approach are empirical values
of mode amplitudes and complex, nonadiabatic parameter $f$. Strictly speaking,
these are amplitudes defined as $\tilde\varepsilon=\varepsilon Y^m_\ell(i,0)$,
where $\varepsilon$ is the intrinsic mode amplitude and $Y^m_\ell(i,0)$
is the spherical harmonic, dependent on the inclination angle, $i$.
Thus, to obtain a value of $\varepsilon$ for nonradial modes one has to know $i$.
In Tab.\,2, we give exemplary values of $|\tilde\varepsilon|$ and $(f_R, f_I)$ for the central model
in the error box assuming two values of the microturbulent velocity, $\xi_t$,
in the atmosphere: 2 and 8 km/s. The last column contains the discriminant $\chi_E^2$.
As one can see, there is a big decrease also in the value of $\chi_E^2$
when atmospheric models with $\xi_t=8$ km/s are used.
\begin{table}
\begin{center}
\caption{Empirical values of $\tilde\varepsilon=\varepsilon Y_\ell^m(i,0)$
and the $f$-parameter for frequencies detected in both photometry and spectroscopy.
Stellar parameters of the central model in the error box (Eq.\,1) were assumed
($M=9.9 M_{\odot},~ \log T_{\rm eff}=4.346, \log L/L_{\odot}=3.94$) and
two values of the microturbulent velocity, $\xi_t$, in stellar atmosphere
were considered: 2 km/s (first lines) and 8 km/s (second lines).}
\begin{tabular}{|c|c|c|c|c|c}
\hline
{frequency} & $|\tilde\varepsilon|$ & $f_R$  &  $f_I$  & $\xi_t$ & $\chi^2$ \\
 &  &  &  &  &  \\
\hline
$\nu_{1}$=5.76328 &   0.0172(7)  &  -9.28$\pm$0.30  &   0.77$\pm$0.30 & 2 & 31.4\\
                  &   0.0174(5)  &  -9.28$\pm$0.20  &   0.75$\pm$0.20 & 8 & 14.3\\
\hline
$\nu_{2}=$5.65388 &   0.0087(3)  &  -9.25$\pm$0.25  &   0.16$\pm$0.25 & 2 & 4.3\\
                  &   0.0088(2)  &  -9.26$\pm$0.19  &   0.15$\pm$0.19 & 8 & 2.3\\
\hline
$\nu_{3}=$5.62002 &   0.0080(4)  &  -9.04$\pm$0.30  &   0.02$\pm$0.30 & 2 & 5.6\\
                  &   0.0081(3)  &  -9.05$\pm$0.27  &   0.02$\pm$0.27 & 8 & 4.4\\
\hline
$\nu_{4}=$5.63725 &   0.0075(2)  &  -9.12$\pm$0.16  &   0.52$\pm$0.16 & 2 & 1.5\\
                  &   0.0075(2)  &  -9.12$\pm$0.15  &   0.50$\pm$0.15 & 8 & 1.3\\
\hline
$\nu_{5}=$7.8982  &   0.0007(1)  & -11.31$\pm$0.80  &   4.55$\pm$0.80 & 2 & 0.3\\
                  &   0.0007(1)  & -11.45$\pm$0.72  &   4.44$\pm$0.71 & 8 & 0.3\\
\hline
$\nu_{6}=$6.2439  &   0.0008(1)  &  -9.09$\pm$1.23  &  -1.28$\pm$1.23 & 2 & 1.3\\
                  &   0.0008(1)  &  -9.16$\pm$1.20  &  -1.27$\pm$1.20 & 8 & 1.2\\
\hline
$\nu_{7}=$6.2629  &   0.0007(1)  &  -7.09$\pm$0.38  &   1.29$\pm$0.38 & 2 & 0.2\\
                  &   0.0007(1)  &  -7.19$\pm$0.41  &   1.26$\pm$0.41 & 8 & 0.2\\
\hline
$\nu_{12}=$6.2236 &   0.0004(1)  &  -4.04$\pm$0.75  &   4.54$\pm$0.75 & 2 & 0.2\\
                  &   0.0004(1)  &  -4.21$\pm$0.73  &   4.43$\pm$0.73 & 8 & 0.2\\
\hline
$\nu_{B}=$0.6144  &   0.0006(2)  &   9.37$\pm$4.14  &  16.60$\pm$4.14 & 2 & 2.1\\
                  &   0.0006(2)  &   9.28$\pm$4.02  &  16.46$\pm$4.02 & 2 & 2.0\\
\hline
\end{tabular}
\label{tab_mse}
\end{center}
\end{table}

Although within linear approximation we cannot compare empirical values
of $\varepsilon$ with their theoretical counterparts, it is interesting
to examine determination of these quantities. The radial mode has by far
the largest intrinsic amplitude; it amounts to about 1.7 \% of the stellar radius.
Components of the $\ell=1,\rm{g_1}$ triplet ($\nu_2,\nu_3, \nu_4$) have comparable values
of $|\tilde\varepsilon|$ between each others but the largest one is obtained for the $\ell=1, m=+1$ mode
and the smallest one, for the centroid mode.
Of course, one has to remember that these quantities have to be divided by $Y_\ell^m(i,0)$.
Consequently, if the star were seen close to the equator-on, then the $\ell=1,m=0$ mode would
have the largest intrinsic amplitude.
Values of $|\tilde\varepsilon|$ for the $\ell=1,\rm{p_1}$ triplet are about ten times smaller
than those for $\ell=1,\rm{g_1}$. The same is true for the only component of the $\ell=1,\rm{p_2}$ triplet
visible in photometry and spectroscopy, i.e., the $\ell=1,m=-1$ mode.
In the case of the SPB mode, $\nu_B=0.614$ c/d, a value of $|\tilde\varepsilon|$
is also comparable with those of the $\rm{p_1}$ triplet.

The second set of unknowns, i.e., the empirical values of the complex parameter $f$,
can be compared with the results of the nonadiabatic pulsation computations.
These quantities will be included in seismic modelling of $\nu$ Eri
and discussed in the next section.
\begin{figure*}
\centering
\includegraphics[width=17.7cm,clip]{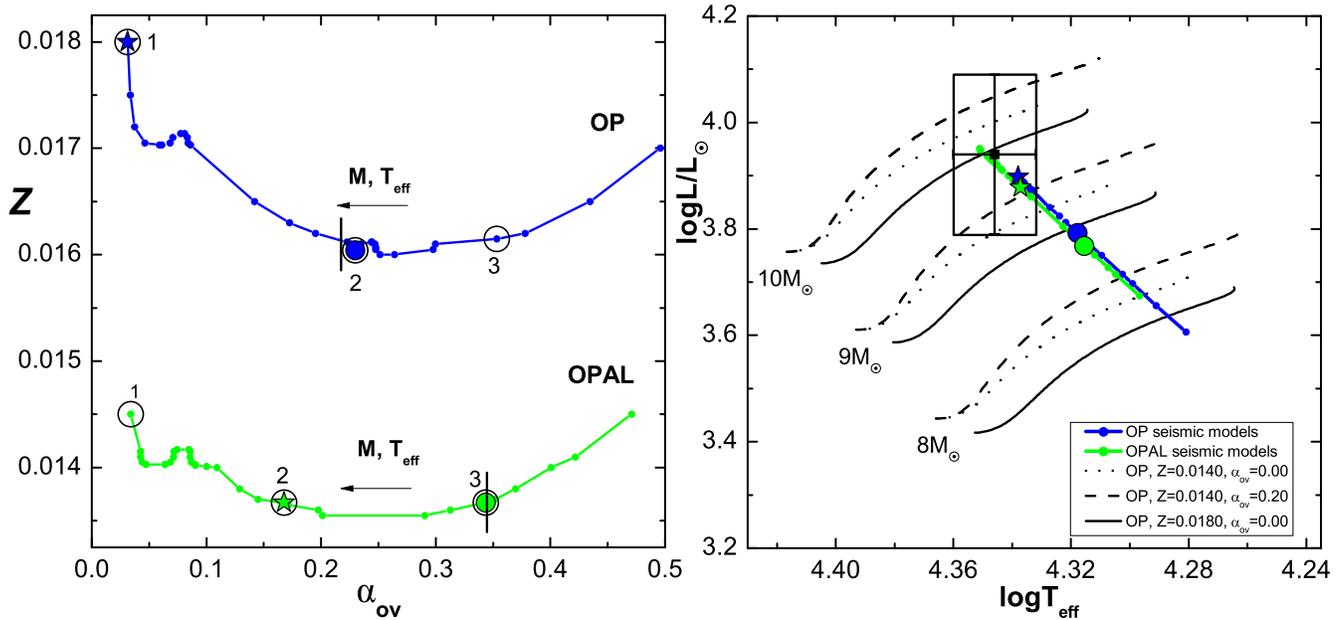}
\caption{On the  left: location of the OP and OPAL seismic models which fit the three centroid frequencies of $\nu$ Eri:
$\nu_1(\ell=0,\rm{p_1}),~\nu_4 (\ell=1,\rm{g_1})$ and $\nu_6(\ell=1,\rm{p_1})$, on the $Z-\alpha_{\rm ov}$ plane.
Models to the left of the vertical tick line have all these three modes unstable.
The big dots mark models fitting also the centroid frequency of the fourth triplet, $\nu_{9}(\ell=1,\rm{p_2})$.
Models marked with the star symbols are used for a comparison of the empirical and theoretical
values of $f$. On the right: position of the OP and OPAL seismic models in the HR diagram
and the observational error box of $\nu$ Eri. Evolutionary tracks are described in the text.
} \label{aaaaaa}
\end{figure*}

\section{Complex astreoseismology}

As in DDW09, our seismic modelling is performed in two stages.
First, we aim at finding stellar models which reproduce observed values of
pulsational frequencies. In the case of $\nu$ Eri, we start by fitting
three centroid frequencies: $\nu_1(\ell=0,{\rm p}_{1})$, $\nu_4(\ell=1,{\rm g}_1)$
and $\nu_6(\ell=1,{\rm p}_1)$. Then, from this family of seismic models
we try to single out those which fit also the forth p-mode frequency,
$\nu_9(\ell=1,{\rm p}_2)$, and the high-order g modes, $\nu_A(\ell=1)$ and $\nu_B(\ell=2)$,
assuming that they are axisymmetric ($m=0$).

In the next step we want to get an agreement between theoretical values of the $f$-parameter
and the empirical ones obtained with the method used already in Sect.\,3 to identify the degrees $\ell$.

All computations were made using the Warsaw-New Jersey evolutionary code
(written in its original version by B. Paczy\'nski) and the pulsational nonadiabatic
code by Dziembowski (1977). We adopt two sources of opacity data:
OP (Seaton 2005) and OPAL (Iglesias \& Rogers, 1996), OPAL equation of state
and the chemical mixture as determined by A04.
At ZAMS, we assumed standard hydrogen abundance of $X=0.7$ and rotational velocity
of 10 km/s. The effect of overshooting from the convective core was included according
to the new formulation of Dziembowski \& Pamyatnykh (2008).
\begin{figure}
\centering
\includegraphics[width=85mm,clip]{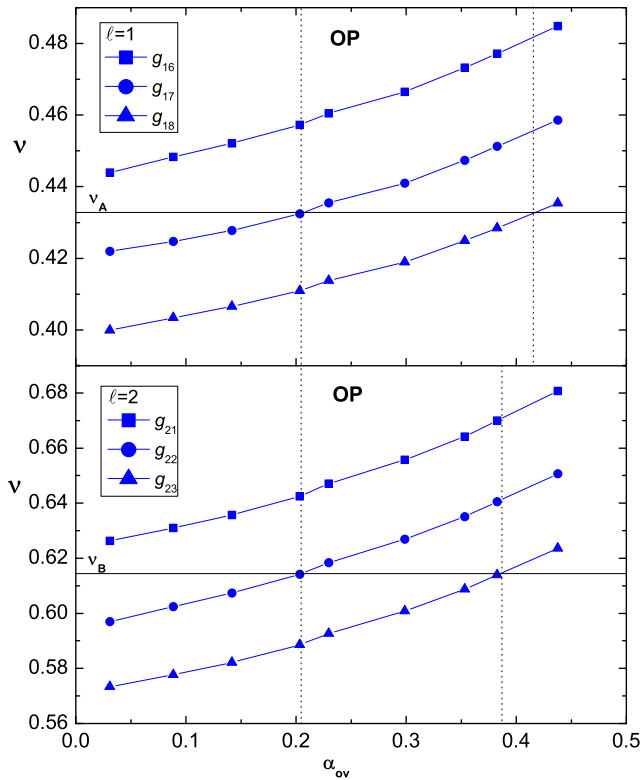}
\caption{Values of frequencies of the OP seismic models around the high-order g modes,
$\nu_A$ and $\nu_B$, identified as $\ell=1$ and $\ell=2$, respectively.
The $\ell=1$ modes correspond to radial orders $n=16,17,18$ and the $\ell=2$ ones
to $n=21,22,23$.} \label{aaaaaa}
\end{figure}

\subsection{Fitting pulsational frequencies}
To find seismic models we survey a wide and dense grid
of stellar parameters: $M,~Z,~T_{\rm eff},~\alpha_{\rm ov}$.
In the left panel of Fig.\,4, we put our seismic models of $\nu$ Eri
found with the OPAL and OP tables on the $Z-\alpha_{\rm ov}$ plane.
All these models reproduce three centroid frequencies: $\nu_1(\ell=0,{\rm p}_1)$,
$\nu_4(\ell=1,{\rm g}_1)$ and $\nu_6(\ell=1,{\rm p}_1)$.
Models to the left of the vertical thick line have all these three modes unstable.
With big dots we indicated models which fit the fourth frequency, $\nu_9(\ell=1,{\rm p}_2)$,
but this mode is stable in these models.
The other symbols (stars and circles) will be discussed later on.
Our accuracy in fitting the frequency $\nu_1$ was about $10^{-7}$ c/d.
For the other two frequencies, $\nu_4$ and $\nu_6$, we achieved an accuracy of $10^{-6} - 10^{-4}$ c/d,
depending on the model, but in most cases it was not worse than $10^{-5}$ c/d.
The accuracy for the fourth frequency, $\nu_{9}(\ell=1,\rm{p_{2}})$ was equal up to $10^{-4}$ c/d.

In the right panel of Fig.\,4 we show location of the OPAL and OP seismic models in the HR diagram
together with the error box of $\nu$ Eri (see Eq.\,1). These models satisfy the following relations:
$${\rm OP}:~~~~~~~~\log L/L_\odot=5.1364\log T_{\rm eff} - 18.3855, \eqno(2a)$$
$${\rm OPAL}:~~~~~\log L/L_\odot=5.0956\log T_{\rm eff} - 18.2214. \eqno(2b)$$
The evolutionary tracks, shown from ZAMS to TAMS, were computed adopting the OP-A04 opacities.
The effect of the heavy elements abundance, $Z$, and the overshooting parameter, $\alpha_{\rm{ov}}$,
on the evolutionary tracks is also presented.
As one can see, models fitting four centroid frequencies are beyond the observational
error box of $\nu$ Eri. The effective temperature is about $2\sigma$ of the observational value.
A similar result was obtained by  Ausseloos et al. (2004), but for much cooler temperatures
or rather unacceptable chemical composition ($X,Z$).

As can be seen from the left panel of Fig.\,4, in general, for $\alpha_{\rm ov}\lesssim 0.25$,
models with larger core overshooting require lower $Z$, smaller masses and effective temperatures
whereas for $\alpha_{\rm ov}\gtrsim 0.25$, models with larger core overshooting require higher $Z$,
smaller masses and effective temperature. We have also found that in some cases for one value of $Z$
there exist several seismic models with different values of core overshooting.
This is connected with the avoided crossing phenomenon.
Results obtained with the OP and OPAL opacities are quite similar but models computed
with the OPAL data require smaller $Z$ by about 0.0025.
The OPAL seismic models have also larger masses and effective temperatures which better match
observational values of $T_{\rm{eff}}$ and $L$ (cf. the right panel of Fig.\,4).
As an example, in Tab.\,3 we give the parameters for selected OP and  OPAL seismic models
marked with circles in the left panel of Fig.\,4.
\begin{table*}
\small
\begin{center}
\caption{Parameters of selected seismic models of $\nu$ Eri (circles in the left panel of the Fig.\,4).
Columns from left to right are: model number, mass, $M/M_\odot$, age, $t$, effective temperature, $\log T_{\rm{eff}}$,
luminosity, $\log{L/L_{\odot}}$, metallicity, $Z$, overshooting parameter, $\alpha_{\rm{ov}}$, radius, $R/R_\odot$,
hydrogen abundance in the centre, $X_{\rm{c}}$, and relative mass of the convective core, $M_c/M$.}
\begin{tabular}{|l|c|c|c|c|c|c|c|c|c|c|c|c|}
\hline
MODEL & $M$ [$M_{\odot}$] & $t$ [My] &$\log{T_{\rm{eff}}}$ & $\log{L/L_{\odot}}$ & $Z$ & $\alpha_{\rm{ov}}$ [$H_{\rm{p}}$] & R [$R_{\odot}$]& $X_{\rm{c}}$&$M_{\rm{c}}/M$ \\
\hline

OP1  &9.6401&17.385&4.3380&3.898&0.01800&0.0312&6.2701&0.24680&0.1847 \\
OP2  &8.5892&23.886&4.3179&3.792&0.01604&0.2297&6.0836&0.26595&0.1965 \\
OP3  &8.0076&29.092&4.3026&3.714&0.01615&0.3531& 5.970&0.27876&0.2018 \\
     &      &      &      &     &       &      &      &       &       \\
OPAL1&9.7706&16.172&4.3510&3.951&0.01450&0.0340&6.2717&0.24691&0.1902 \\
OPAL2&9.0546&19.935&4.3372&3.879&0.01367&0.1677&6.1503&0.26078&0.1980 \\
OPAL3&8.1709&26.325&4.3155&3.768&0.01367&0.3436&5.9837&0.27933&0.2058 \\
\hline
\end{tabular}
\medskip
\label{char_models}
\end{center}
\end{table*}
\begin{figure}
\centering
\includegraphics[width=85mm,clip]{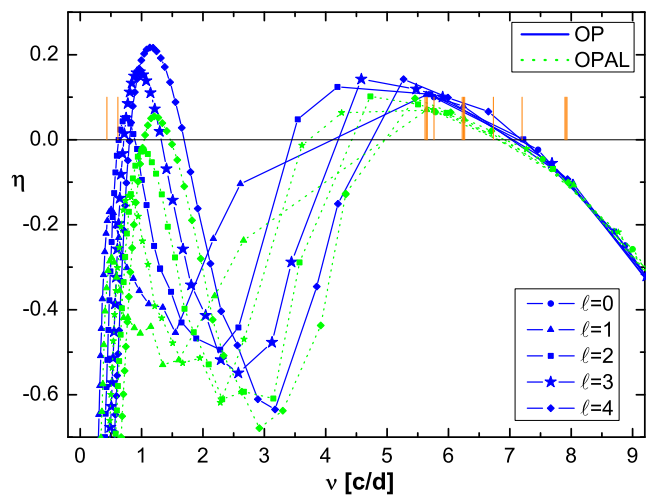}
\caption{The normalized instability parameter, $\eta$, plotted against the mode frequency with degrees $\ell=0-4$.
Two seismic models with about the same effective temperature were considered: OP1 and OPAL2 (cf. Tab.\,3).
} \label{aaaaaa}
\end{figure}

The radial orders for the SPB modes, estimated from our seismic models,
are $n\approx16-18$ for $\nu_A=0.433$ c/d and $n\approx21-23$ for $\nu_B=0.614$ c/d.
Let us assume that these modes are axisymmetric $(m=0)$ and check whether there is
a seismic model amongst those presented in Fig.\,4 which reproduces also these high-order g modes.
In Fig.\,5, we show the frequency values of these orders for the OP seismic models
as a function of the overshooting parameter $\alpha_{\rm ov}$.
In the top and bottom panel we plot frequencies around $\nu_A$ and $\nu_B$, respectively.
We can see that in each case there are two seismic models which reproduce
these high-order g modes. The accuracy of these fits amounts to $10^{-4}$ c/d.
The corresponding radial orders for $\nu_A$ are $n=17$ and $n=18$
at $\alpha_{\rm ov}\approx0.20$ and 0.42, respectively. In the case of $\nu_B$,
we get $n=22$ and $n=23$ at $\alpha_{\rm ov}\approx0.20$ and 0.39, respectively.
We can see that the OP seismic model with $\alpha_{\rm ov}\approx0.20$ fits simultaneously $\nu_A$ and $\nu_B$.
The next coincidence is that this is the model close to the OP2 model (see Tab.\,3)
which fits also the forth pressure mode, $\nu_9(\ell=1,\rm{p_2})$.
A similar result has been obtained for the OPAL seismic models and with about the same values
of $\alpha_{\rm ov}$,  but none of them coincides with the OPAL3 model
which fits $\nu_9(\ell=1,\rm{p_2})$.

Now, let us come back to the instability of the modes.
In Fig.\,6 we plot the normalized instability parameter, $\eta$,
as a function of frequency for two seismic models with about the same effective temperature
but differing in all other parameters and the opacity data (models OP1 and OPAL2 in Tab.\,3).
These seismic models are marked with star symbols in Fig.\,4.
Let us remaind the reader that a mode is unstable if $\eta>0$.
As we can see, the OP seismic models are more unstable; this results mainly
from higher metallicity for the OP1 model ($Z=0.018$). In general, at the same $Z$,
pulsation computations with the OP tables give more instability for low frequency modes (the SPB modes)
whereas with the OPAL data we get a wider instability in the high frequency range (the $\beta$ Cep modes).
This is because the $Z-$bump in the OP tables is located in deeper layers
where the thermal time scale is longer.
As we have already mentioned, the three highest amplitude centroid frequencies ($\nu_1,~\nu_4,~\nu_6$)
are excited in most seismic models (to the left of the vertical thick line in the left panel of Fig.\,4).
In all our seismic models the frequency $\nu_9$, corresponding to the $\ell=1,p_2$ mode,
is stable. The highest values of $\eta$ for this mode are $\eta=-0.10$ in the OP models
and $\eta=-0.04$ in the OPAL models.
As for the high-order g modes, the frequency $\nu_A$, identified as the $\ell=1$ mode,
is stable in all seismic models, but $\nu_B$, which is an $\ell= 2$ mode, reaches instability very often.

\subsection{Constraints on opacities from the $f$-parameter}
Empirical values of $f$ were obtained from the method of simultaneous
determination of $f$ and $\ell$ (cf. Sect.\,3).
\begin{figure*}
\centering
\includegraphics[width=17.7cm,clip]{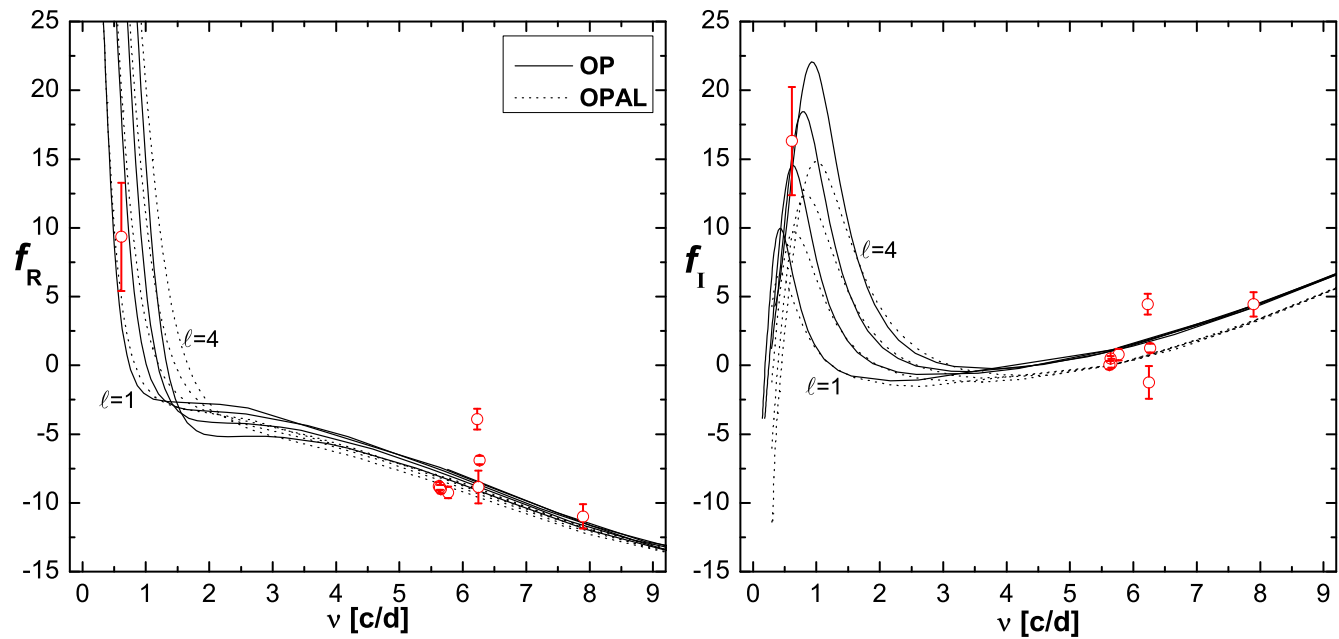}
\caption{Comparison of the empirical and theoretical values of $f$ in the whole range of pulsation frequencies
observed in both photometry and spectroscopy of $\nu$ Eri. The real and imaginary
parts of $f$ are plotted in the right and left panel, respectively. Theoretical values
corresponding to the OP1 and OPAL2 seismic models are marked with star symbols in Fig.\,4 (cf. also Tab.\,3).
Modes with the degree $\ell$ from 0 to 4 were considered.}
\label{aaaaaa}
\end{figure*}
In Fig.\,7 we compare the empirical and theoretical values of $f$ for all nine frequencies
of $\nu$ Eri found in both photometry and spectroscopy. Theoretical values of $f$ corresponding to
the OP1 and OPAL2 seismic models in Tab.\,3 (marked also with star symbols in Fig.\,4)
are plotted as solid and dotted lines, respectively. Modes with the spherical harmonic
degrees from $\ell=0$ to 4 were considered.
As one can see, there is quite a good agreement for the most pulsational frequencies
including the SPB-type mode, $\nu_B=0.614$ c/d, which was identified as $\ell=2$.
Unfortunately, the frequency $\nu_A=0.433$ c/d is not visible in spectroscopy.
Around $5.6-5.8$ c/d we have four points which overlap, because frequencies
$\nu_1$ and $\nu_2,\nu_3,\nu_4$ are very close to each other.
The discrepancy for the remaining frequencies is caused by inadequate accuracy of the photometric
amplitudes and phases in the case of the low-amplitude modes, e.g., $\nu_{12}$.

Having empirical values of $f$ in such a wide range of frequency is a big advantage.
Particularly because the dependence of $f$ on pulsational frequency and the mode degree, $\ell$,
is different for p and g modes. In general, the $f$-parameter is determined
by pulsational frequency and the shape of eigenfunction as
$$f\sim \left( \frac{\ell(\ell+1)}{\sigma^2} -\sigma^2-4 \right), \eqno(3)$$
where $\sigma=\omega \sqrt{R^3/(GM)}$ is the dimensionless frequency, $\omega$ is
the angular frequency, $\omega=2\pi\nu$, and $G,~M,~R$ have their usual meaning.
Thus, for p modes the second term dominates and $f$ is independent of $\ell$.
If frequencies are low (g modes), the first term dominates and
we have a strong dependence of the $f$-parameter on $\ell$.
This is clearly visible in Fig.\,7.
We can see also that while values of $f$ for the p modes change rather slowly
with frequency, a rapid variation of $f$ occurs for the g modes.

The best accuracy of the values of $f$ was obtained for the radial mode
and the three $\ell=1, \rm{g_1}$ modes. Their frequencies are very
close to each other and so are their $f$-parameters (cf. Tab.\,2).
Therefore, further consideration will concern the $f$-parameter of the radial mode, $\nu_1$.
In Fig.\,8 we show a comparison of the empirical and theoretical values of $f$
of the OP and OPAL seismic models.
We consider models with effective temperatures within $3\sigma$ of the observational value.
As can be seen from Fig.\,8, only with the OPAL tables we can get an agreement of the $f$-values
within the errors and this is achieved for lower values of $\log T_{\rm eff}$ and lower metallicity.
The same result was obtained for the $\beta$ Cep star $\theta$ Oph (DDW09).
The OPAL model which fits four $\nu$ Eri frequencies has the $f$-parameter
marginally consistent with our empirical determination (the big dot).
We do not know whether this result is a stroke of good luck or has some physical meaning.
The effect of the microturbulent velocity in the atmosphere, $\xi_t$,
on the values of $f$ is negligible (cf. Tab.\,2).
We did not examine the effect of the hydrogen abundance, $X$,
because, as has been shown by DDW09, in the case of early B-type pulsators,
the $f$-parameter is only weakly sensitive to $X$.
\begin{figure}
\centering
\includegraphics[width=86mm,clip]{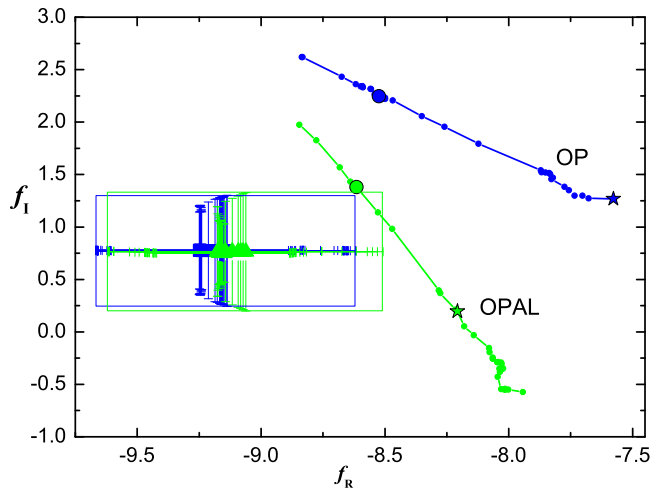}
\caption{Comparison of the empirical values of $f$ for the radial
mode, $\nu_1(\ell=0,\rm{p_1})$, with theoretical ones corresponding
to the OP and OPAL seismic models of $\nu$ Eri.} \label{aaaaaa}
\end{figure}
\begin{figure}
\centering
\includegraphics[width=84mm,clip]{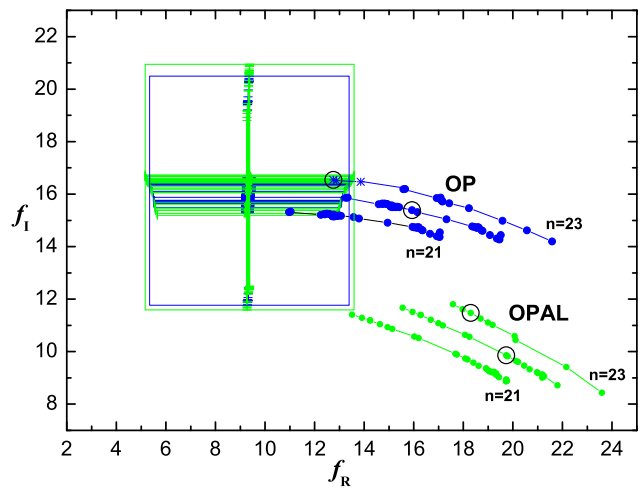}
\caption{Comparison of the empirical values of $f$ for the g, $\ell=2$ mode, $\nu_B$,
with theoretical ones corresponding to the OP and OPAL seismic models of $\nu$ Eri.
The same radial orders, $n$, as in Fig.\,5 were considered. Circles mark models
which fit the frequency $\nu_B$. Two OP models with the radial order $n=23$
are outside the $3\sigma$ error of the effective temperature, marked
with asterisks.} \label{aaaaaa}
\end{figure}

As we have mentioned above, it would be of particular interest to compare the theoretical and empirical
values of $f$ also for the high-order g modes. Here we can do this for the frequency $\nu_B$
which was detected in both photometry and spectroscopy.
In Fig.\,9, we show such a comparison using the same seismic models as those in Fig.\,8.
With circles we marked models fitting the frequency $\nu_B$ and with asterisks,
three OP models with the radial order $n=23$ which are outside the $3\sigma$ error
of the effective temperature.
The result is quite intriguing.  While in the case of the $\beta$ Cep mode
the empirical values of $f$ agree with the OPAL computations,
for the SPB mode, $\nu_B$, an agreement for both real and imaginary part
of $f$ can be achieved with the OP data.

Moreover, one can notice that in the case of both, the p modes
and the high-order g mode, theoretical values of $f$ of the OPAL and OP models agree better
in the real parts than in the imaginary ones.

Here, we have to discuss also a possibility that the frequency $\nu_B$ is a non-axisymmetric mode.
As we mentioned in Sect. 2, the rotational velocity of $\nu$ Eri is very slow
and the effect of the azimuthal order, $m$, on $f$ can be entirely neglected.
As for the frequency dependence of $f$ our justification is as follows.
Taking the stellar radius of $R\approx6R_{\odot}$, the rotational frequency
is $\nu_{\rm rot}\approx 0.019$ c/d. Given the high-order g modes,
the Ledoux constant can be approximated by $C_{n\ell}\approx\frac1{\ell(\ell+1)}$.
Thus for $\ell=2$ we get the rotational splitting of $\frac 56\nu_{\rm rot}\approx 0.016$ c/d.
In so narrow frequency range, i.e., $0.614\pm 0.016$ c/d, both the real and imaginary parts of $f$
are nearly constant. Therefore, although the assumption that $\nu_B$ is the $m=0$ mode
can be invalid, it does not change above results.

\section{Conclusions}

We presented comprehensive asteroseimic modelling of the $\beta$ Cep/SPB star $\nu$ Eridani.
We started from mode identification for all fourteen pulsation frequencies detected
in the light variations from the last photometric campaign (J05).
For nine frequencies, seen also in the radial velocity variation, we were able to apply
the method of simultaneous determination of the degree $\ell$ and the nonadiabatic parameter $f$ (DD05).
We confirmed identifications of $\ell$ from previous determinations. Moreover,
mode degrees of the SPB frequencies were unambiguously determined.
We got $\ell=1$ for $\nu_A=0.433$ c/d and $\ell=2$ for $\nu_B=0.614$ c/d.

In the next step, we looked for stellar models which fit the three centroid frequencies:
$\nu_{1}(\ell=0,\rm{p_{1}})$, $\nu_{4}(\ell=1,\rm{g_{1}})$ and $\nu_{6}(\ell=1,\rm{p_{1}})$.
Using opacities from the OP and OPAL tables and the A04 chemical mixture,
we found a family of seismic models with different masses, effective temperatures,
metallicities and core overshooting parameters. The OPAL seismic models agreed better
with observational values of the effective temperature, $T_{\rm{eff}}$, and luminosity, $L$.
With two opacity data, we managed to find also models which fit the fourth centroid
frequency, $\nu_{9}(\ell=1,\rm{p_{2}})$, but with worse accuracy in comparison with
the first three ones; these models have effective temperatures about $2\sigma$ cooler
than the observational value.

Moreover, it turned out that some of our seismic models reproduce simultaneously also
the high-order gravity modes, $\nu_A$ and $\nu_B$. This was the case for models with
the overshooting parameter, $\alpha_{\rm ov}$, of about 0.2 and 0.4, independently of
which opacity data were used. Additionally, the OP seismic model with $\alpha_{\rm ov}=0.2$
is close to the one which fits also the fourth p mode, $\nu_9$.

Then we extended our seismic analysis by adding the requirement of fitting
the $f$-parameter. A comparison of the empirical and theoretical values of $f$
for nine frequencies, which appeared in both photometry and spectroscopy,
showed an overall agreement, especially for the modes which have the
photometric amplitudes and phases determined with high accuracy.
These nine frequencies are in the range of $\nu\in(0.6,8)$ c/d, encompassing eight p modes
and one high-order g mode. A very promising result is that the empirical value of $f$
for the SPB-type mode, $\nu_B=0.6144$ c/d, agrees with theoretical computations.

A more detailed analysis of the $f$-parameter for the radial fundamental mode, $\nu_1$,
showed that a concordance could be achieved only with the OPAL opacity data.
The same result was obtained by us for the $\beta$ Cep variable $\theta$ Oph (DDW09).
On the other hand, the same comparison for the high-order g mode, $\nu_B$, favoured the OP data.
This discrepancy calls for an explanation and indicates that
something is still missing in the computations of opacity data.
Complex asteroseismology of hybrid B-type pulsators can bring us closer to a solution.

\section*{Acknowledgments}
We thank Miko{\l}aj Jerzykiewicz for the light curves of $\nu$ Eri
in the Str\"omgren passbands and for carefully reading the manuscript.
We are grateful also to the reviewer, Gerald Handler, for his comments.
The work was supported by the Polish MNiSW grant N N203 379636 and
the European Helio- and Asteroseismology Network (HELAS), No. 026138.

\label{lastpage}

\end{document}